\documentclass[12pt]{article}
\usepackage{graphicx}
\usepackage{color}
\usepackage{amsmath,amssymb}
\usepackage{epsfig}
\usepackage{epstopdf}
\usepackage{hyperref}
\begin{document}

\begin{titlepage}
\begin{center}
{\hbox to\hsize{arXiv:1512.04338 [quant-ph]  \hfill  IZTECH-P06/2015}} 

\bigskip
\vspace{3\baselineskip}

{\Large \bf Statistical Approach to Tunneling Time in Attosecond Experiments}

\bigskip

\bigskip

{\bf Durmu{\c s} Demir}\\
\smallskip

{ \small \it  
Department of Physics,
{\.I}zmir Institute of Technology,
{\.I}zmir, TR35430, Turkey}

\smallskip
{\bf Tu{\~g}rul G{\"u}ner}\\
\smallskip
{\small \it Department of Material Science and Engineering, 
{\.I}zmir Institute of Technology,
{\.I}zmir, TR35430, Turkey}

\bigskip

{\tt  demir@physics.iztech.edu.tr} \\
{\tt tugrulguner@iyte.edu.tr}

\bigskip

\vspace*{.5cm}

{\bf Abstract}\\
\end{center}
\noindent
Tunneling, transport of particles through classically forbidden
regions, is a pure quantum phenomenon. It governs numerous phenomena
ranging from single-molecule electronics to donor-acceptor
transition reactions. The main problem is the absence of a universal
method to compute tunneling time. This problem has been attacked in
various ways in the literature. Here, in the present work, we show
that a statistical approach to the problem, motivated by the
imaginary nature of time in the forbidden regions, lead to a novel
tunneling time formula which is real and subluminal (in contrast to
various known time definitions implying superluminal tunneling). This entropic tunneling time, as we call it, shows good agreement  with the tunneling time measurements in laser-driven He ionization. Moreover, it sets an accurate range for long-range electron transfer reactions. The entropic tunneling time is general enough to extend to the photon and phonon tunneling
phenomena.

\bigskip

\bigskip

\end{titlepage}

\section{Introduction}

Tunneling, transport of subatomic particles through the regions of
space forbidden to classical motion, is a pure quantum phenomenon.
Its physical relevance was first established by Gamow in his
analysis of the $\alpha$-decay \cite{gamow,ilkler,ilkler2}. The
tunnel diode \cite{esaki,esaki2} of Esaki was its first
technological application. Undoubtedly, scanning tunneling
microscope (STM) \cite{stm} of Binning and Rohrer started a new pace
in scientific and technological advancements. Today, for example, it
is known that electron transfer reactions involve tunneling as the
underlying mechanism. It governs acceptor-donor transition processes
so that charge separation by electron transfer reaction takes place
at photosynthetic reaction centres\cite{reactcent} after excited
electrons are transferred by antenna pigments in consequence of the
coherent electron energy transfer\cite{antenna} during the
photosynthesis. Tunneling is a ubiquitous mechanism that underlies
numerous physical \cite{phys}, chemical \cite{chem}, biological
\cite{bio} and technological phenomena \cite{roy}.

Tunneling time, the time elapsed during the tunneling process, is
crucial for determining reaction speeds of tunneling-enabled rare
processes ranging from high-speed electronic devices to nuclear
fusion. Moreover, it also plays an important role while determining
the electron transfer reaction rates based on the interaction of
electron with the corresponding vibrational mode of the
molecule\cite{electrontransfer}. Recently, transition from region
where Born-Oppenheimer approximation holds, to region where it
breaks down is also investigated in terms of tunneling
time\cite{BOapprox}. In fact, with the advent of strong laser
ionization experiments \cite{exp0, exp1,exp11,exp12,exp13}, it is
becoming possible to measure the tunneling time \cite{exp2,exp2-new}
where certain metrological problems \cite{exp-problem1,
exp-problem2} with the detection of the tunneling particle are shown
to be surmountable \cite{exp-resolve1, exp-resolve2}. Strong laser
fields enable electrons to tunnel out of atoms, where the potential
barrier formed forms a testbed for models of tunneling time
\cite{keldysh,keldysh2}.

Tunneling time depends on what kinetic theory is set forth for
tunneling process, and therefore, the literature consists of various
time definitions
\cite{time-review,time-review2,time-review3,time-review4}. They
include traversal time through modulated barriers
\cite{buttiker-landauer, buttiker-landauer2, buttiker-landauer3,
japan-rec}, spin precession time \cite{larmour, larmour4,larmour5},
flux-flux correlation duration \cite{pollak-miller}, phase
stationary time \cite{wigner,wigner2,wigner3}, polymer approach \cite{demir-sargin}, and Feynman path
integral (FPI) averaging of the classical time
\cite{times-are-averages,times-are-averages2,times-are-averages3}.
Some of them are complex, some are difficult to associate with
tunneling and some suffer from superluminality. Interestingly,
contrary to their raison d'etre, all these tunneling times utilize a
sort of time operator since they involve derivatives with respect to
energy or potential. This is not the case for FPI averaging yet the
resulting time is still controversial because classical trajectories
live in imaginary time and their probability amplitudes interfere
\cite{interference, interference2, interference3, interference4,
FPI-method}. At present, the problem with these and other tunneling
time definitions is that they seem incapable of explaining the
experimental data as was comparatively analysed and experimented in
\cite{exp-incele}. In view of the growing scientific and
technological needs, however, it is necessary to have a working
model that can reliably estimate the tunneling time for a given
potential barrier.

The present work reports on a novel formulation of the tunneling
time. The formulation, based on a statistical description of the
evanescing particle in the classically forbidden region, gives a
tunneling time which shows {good agreement with the experimental data} compared to all the
widely-used time definitions. The essence of the formulation is
that, in the classically forbidden region time flows in imaginary
direction, and correspondence between imaginary time in quantum
mechanics and temperature in statistical mechanics enables a
statistical formulation for tunneling. The resulting thermal energy,
through the uncertainty principle, sets a generalised time interval
depending on the transmission amplitude. Next, this model is applied
numerically to the laser-driven He ionization covering recent
experimental data \cite{exp-incele} and to the electron transfer
reactions. It is found that entropic tunneling time is in good
agreement with the experimental data for the former, and sets the
validity range of electron transitions in the long range electron
transfer reactions, for the latter.

The energy-time uncertainty, which we utilize for the thermal energy
of the tunneling particle, has been utilized in a different
tunneling time study \cite{uncertain}. In that related work, it is
assumed that exchange between the kinetic energy of the electron and
the potential energy describing the tunneling region leads to an
uncertainty in the total energy of the electron. It is then taken
that the uncertainty in this total energy is proportional to the
potential energy at the exit point of the electron $\Delta E \propto
|V_{exit}|$. The model potentials used in \cite{uncertain}, which are based
on effective nuclear charge models described already in Section 4.1, are also used in this study to compute the tunneling time
of the experiment. As a result, even though these two models have
not much in common other than the energy-time uncertainty, results
given here in Section 4.1 and in \cite{uncertain} show good
agreement for He ionization in attosecond experiments. It is with
further experimental data that possible relationship between our
approach and that of \cite{uncertain} may be settled.

In Sec. 2 below given are derivation of the entropy characterizing
the tunneling particle and definition of the tunneling time. Sec. 3
compares the entropic tunneling time with other times. Sec. 4 is devoted to numerical study including laser-induced He ionization
and electron transfer reactions in two separate subsections. Sec. 5
concludes the work and gives future prospects on applications to
different tunneling-enabled phenomena and extensions to photon and
phonon tunnelings.

\section{Entropic Tunneling Time}
\begin{figure}
\includegraphics[scale=.5]{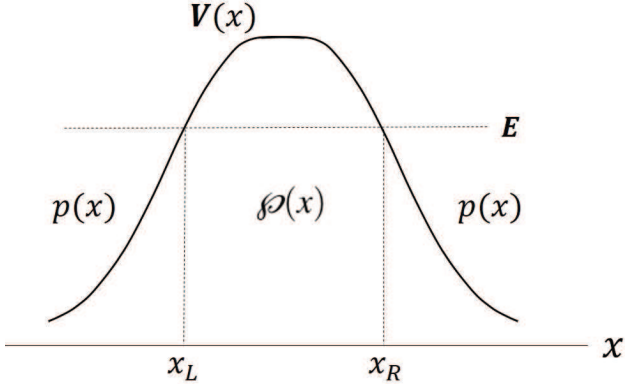}
\caption{{Schematic illustration of the tunneling setup. Potential is smooth. Its central region is the tunneling region.}}
\label{tunnelscheme}
\end{figure}

In classical dynamics, time elapsed while a particle moves from one
point to another cannot be determined without knowing its momentum
at each point in between. This is because momentum is generator of
the translation and, with strict energy conservation, it becomes
${\displaystyle{p(x)=\sqrt{2 m \left( E - V(x)\right)}}}$ for a
particle moving along $x$ axis with mass $m$, potential energy
$V(x)$ and total energy $E$. The particle turns back to its region
of incidence from the turning points $x_L$ and $x_R$ at which
$E=V(x_L)=V(x_R)$. The setup is illustrated in Fig.
\ref{tunnelscheme}. In this forbidden region (the central region in
the figure) $E<V(x)$ and classical dynamics proceeds with imaginary
momentum $p(x) = i \wp(x)$ with
\begin{eqnarray}
\label{c-mom}
\wp(x) = \sqrt{2 m \left(V(x) - E\right)}
\end{eqnarray}
and the lapse time it defines becomes also imaginary $t \rightarrow -i \tau_c$ (see \cite{time} for time arrow) with
\begin{eqnarray}
\label{dc-time}
\tau_c = \int_{x_L}^{x_R} \frac{m dx}{\wp(x)}
\end{eqnarray}
defining what one may call the classical tunneling time. Its
imaginary nature implies that traversing the classically forbidden
region costs no real time. The scattering process seems instantaneous and
acausal. There is an ongoing debate \cite{exp-problem1,exp-problem2,exp-problem3}
on the nature of the tunneling time. In the present work, assumption is that
tunneling time is real and finite \cite{measurable,measurable2,measurable3}
and is consistent with earlier discussions in \cite{real-time,real-time2,real-time3}. \\

By definition, passage of the particle from $x_L$ to $x_R$ with
conserved energy ($E<V(x)$) defines quantum tunneling. It is a pure
quantum phenomenon. {Tunneling time, however, is not a quantum
concept \cite{pauli}. The reason is that time is not an observable
representable by some operator as otherwise it would stop flowing in
its eigenstates. Physically correct description of tunneling time, a
deterministic dimension, might therefore involve an amalgamate of
the classical description above and the quantum behaviour. (Despite
these, recently  Bauer proposed a self-adjoint time operator based
on Dirac's formulation of relativistic quantum mechanics
\cite{bauer1,bauer2}. This proposed time is correlated with
zitterbewegung type fluctuations, and has been claimed \cite{bauer3}
to agree the experiment \cite{exp-incele}. Loss of probability
interpretation in relativistic realm and averaging-out of the
zitterbewegung term over positive-energy states make this time
definition curious.) To this end, one first notes that time flow is
directly correlated with particle's momentum (as defined in
Eq.(\ref{dc-time})), and thus, tunneling time must be addressed in
momentum eigenstates $\psi_m(x)$, not in energy eigenstates
$\psi_e(x)$ (Schrodinger equation refers to energy not
displacement). The second point is that, as a means of ensuring
penetration of the particle into the classically forbidden region,
classical momentum must function as the momentum eigenvalue
associated with $\psi_m(x)$. More precisely, $\psi_{m}(x)$ must
satisfy the eigenvalue equation
$\hat{p}\psi_m(x)=\sqrt{2m(E-V(x))}\psi_m(x)$ where
$\hat{p}=-i\hbar\frac{d}{dx}$ is the momentum operator. In tunneling
region, where $p(x)=i\wp(x)$, one gets
\begin{equation}
\label{mom-eq}
\begin{split}
\frac{d}{d x} \psi_m(x) =- \frac{\sqrt{2m(E-V(x))}}{i \hbar} \psi_m(x) \equiv -\frac{\wp(x)}{\hbar} \psi_m(x) \\
\end{split}
\end{equation}
in agreement with energy conservation. This momentum eigenvalue
shows that particle's momentum inside the tunneling region is
strongly correlated with the potential function. At every point
under the barrier, particle's momentum changes point to point and
leads to the classical time in (\ref{dc-time}) integrated over the
history of motion. This equation can always be integrated to find
\begin{eqnarray}
\psi_m(x) = \psi_m(x_L) \exp\left\{-\frac{1}{\hbar} \int_{x_L}^x \wp(\tilde{x}) d\tilde{x} \right\}
\end{eqnarray}
which is an evanescent wave that decays exponentially as the
particle penetrates farther and farther from $x_L$. This evanescent
behaviour is the key aspect of the tunneling phenomenon. It encodes
all the essential ingredients needed to describe the tunneling
dynamics. To ensure that particle is in the tunneling region, the
probability to find the particle at $x$ satisfying $x_L \leq x \leq
x_R$ (proportional to $\psi_m^{\dagger}(x)\psi_m(x)$) can be
normalized with respect to the probability that it got into the
tunneling region at $x_L$ (proportional to
$\psi_m^{\dagger}(x_L)\psi_m(x_L)$). This way, probability to find
the particle at $x=x_R$ becomes
\begin{equation}
\label{probability}
\begin{split}
p_m = \frac{\psi_{m}^{\dagger}(x_R) \psi_{m}(x_R)}{\psi_{m}^{\dagger}(x_L)\psi_{m}(x_L)} = \exp\left\{-\frac{2}{\hbar}\int_{x_L}^{x_R} \wp(x) dx \right\} \equiv e^{-2 \Phi} \\
\end{split}
\end{equation}
where, for future use, one introduces the dimensionless quantity
\begin{eqnarray}
\label{phase}
\Phi = \frac{1}{\hbar} \int_{x_L}^{x_R} \wp(x) dx
\end{eqnarray}
which measures action of the particle in units of $\hbar$. It now
becomes clear that $p_m$ vanishes for infinitely wide and infinitely
high potential barriers ($p_m\rightarrow 0$ as $\Phi \rightarrow
\infty$) and equals unity if the barrier is absent ($p_m\rightarrow
1$ as $\Phi \rightarrow 0$). Nevertheless, pertaining to a definite
momentum state, it cannot tell whether tunneling has really been
completed or not. The question of whether the particle has tunneled
or reflected is determined by the tunneling transmission probability
$p_{t}$ not $p_m$. It is obtained by solving the Schroedinger
equation
\begin{equation}
\label{en-eq}
\begin{split}
\frac{d^2}{dx^2} \psi_e(x) =-\frac{2m}{\hbar^2}(E-V(x))\psi_e(x) \equiv \left(\frac{\wp(x)}{\hbar}\right)^2 \psi_e(x) \\
\end{split}
\end{equation}
wherein the energy eigenfunction $\psi_{e}(x)$, unlike the momentum
eigenfunction $\psi_{m}(x)$, involves both right-evanescing and
left-evanescing waves. They give rise to the well-known transmission
and reflection probabilities \cite{ilkler2,phys,roy}.

In general, when interpreted as inverse temperature, the imaginary
time is known to transform propagators in quantum mechanics into
partition functions in statistical mechanics \cite{feynman-hibbs}.
This ensures that tunneling time can be addressed in a statistical
framework despite the peculiarity that what is referred to here is a
single particle (not a collection of particles as in the usual
statistical thermodynamics). Moreover, the particle is not in a
mixed state (it is the evanescent wave decaying towards $x_R$). This
means that entropy, required by temperature, must be defined in a
different way. Fundamentally, entropy is given by logarithm of the
number of microstates. The requisite microstates can be identified
by excogitating to volume of the particle's phase space in units of
$\hbar$. This quantity is precisely the main variable $\Phi$ and
counting it means essentially the Bohr-Sommerfeld quantization rule.
It refers to periodic dynamics which is what happens in tunneling
region when potential is effectively inverted with imaginary time.
Fig. \ref{phasespace} gives an illustration of how $\Phi$ can be
related to the microstates pertaining to a tunneling particle.
Identification of $\Phi$ with the number of microstates enables a
statistical description of tunneling dynamics. For a correct
formulation, one observes that entropy, specific to the tunneling
region, must vanish when the barrier is absent ($\Phi \rightarrow
0$). This means that the number of microstates can be taken as $2
\Phi +1$, where the factor of 2 is put for getting an integer since
$\Phi$ is half-integer and the 1 is added for obtaining correct
limiting value as $\Phi \rightarrow 0$. As a result, the tunneling
particle acquires entropy \cite{neden} $k_B\, p_m \log\left(1 +
2\Phi \right)$ which equals, after relating $\Phi$ to $p_m$ via
(\ref{probability}), the compact expression

\begin{figure}
\includegraphics[scale=.5]{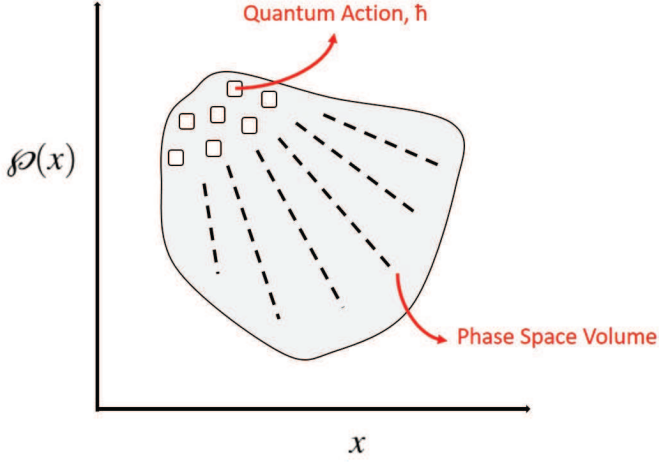}
\caption{{Schematic representation of the phase space volume and total quantum actions}}
\label{phasespace}
\end{figure}

\begin{eqnarray}
\label{entropy}
S(p_{m}) = k_B\, p_m \log\left(1 - \log p_m\right)
\end{eqnarray}
which satisfies $S(p_m=0) = 0$, $S(p_m=1)=0$ and $S(p_m) \geq 0$.
Needless to say, this loglog structure is an artifact of the
exponential relationship between $\Phi$ and $p_m$ in equation
(\ref{probability}). To see this, one notes that use of the uniform
probability $p_{u}= 1/(1+2\Phi)$ would result in the familiar
Boltzmann entropy $-k_B p_{u} \log p_{u}$ in place of the loglog
entropy in (\ref{entropy}). In consequence, the entropy formula
(\ref{entropy}) is specific to the tunneling region and gives a
statistical description of its imaginary-time and evanescent
dynamics.

In this statistical formulation, the energy $E$ of the particle
functions as the total internal energy. The thermal energy, on the
other hand, equals the entropy rate of change of the internal
energy. This is so because tunneling involves a single particle with
quantum probabilistic qualities, and the energy rate of change of
the tunneling entropy in (\ref{entropy}) gives
\begin{equation}
\label{temperature}
\begin{split}
\frac{1}{{k_B T}} \equiv \frac{1}{k_B} \frac{\partial S(p_m)}{\partial E} = - \frac{2 \tau_c}{\hbar}\, e^{-2\Phi} \left[\frac{1}{1+2\Phi} + \log \frac{1}{1+2\Phi}\right]
\end{split}
\end{equation}
where $\tau_c$ is the classical tunneling time in (\ref{dc-time})
and is $\tau_c=-\hbar \partial \Phi / \partial E$. It is not
surprising that the temperature $T$ is proportional to the
reciprocal of $\tau_c$ \cite{feynman-hibbs}. This temperature,
defined for a single particle owing to its quantum indeterminacy,
involves both the Boltzmann constant $k_B$ and Planck constant
$\hbar$. In a true thermodynamical system there can exist no $\hbar$
sensitivity in the classical limit. The specialty of tunneling is
that it is a pure quantum phenomenon having no classical limit.
Thus, the statistical description of tunneling we are presenting
necessarily involves both $k_B$ and $\hbar$.

In quantum tunneling, particle's energy $E$ stays constant
throughout the barrier, and particle ionizes to continuum with the
same energy E at the end of tunneling.(In experimental environments,
time measurement after ionization can alter energy $E$. Such effects
do not influence tunneling time definitions as they all refer to the
barrier region. This is valid also for the ETT.) In contrast to $E$,
however, the thermal energy $k_B T$ varies with barrier shape as in
(\ref{temperature}) and, physically, it sets a finite time interval
$\Delta t$ in the philosophy of the energy-time uncertainty product.
This time interval, as insured by construction of the probability
$p_m$ in (\ref{probability}), must be nothing but the time elapsed
while the particle gets from $x_L$ to $x_R$. One here notes that
these turning points vary with $E$ and, in general, lower the $E$
larger the $x_R - x_L$ and longer the tunneling duration. Its dependence on $p_m$ suggests that $\Delta t$ can be related to tunneling duration. In this sense, let us assume that this is some sort of a tunnelling time, and study the consequences of this assumption. Thus, one defines tunneling time as
\begin{eqnarray}
\label{uncertain}
\Delta t = \frac{\hbar}{2 \Delta E_{ther}}
\end{eqnarray}
which is no different than the energy-time uncertainty relation. In
here, $\Delta E_{ther}$ is the thermal energy needed for completing
the tunneling, and it can be written as
\begin{eqnarray}
\label{deltaE}
\Delta E_{ther} = p_{t} \left( 2 \pi k_B T \right)
\end{eqnarray}
in view of the finite-temperature quantum fluctuations. Here,
$p_{t}$ is the tunneling transmission probability (computed from
solution $\psi_e$ of the Schroedinger equation). The thermal energy
$2\pi k_B T$ might be interpreted as the splitting between Matsubara
levels \cite{matsubara,matsubara2} in finite-temperature quantum
theory. At last, the tunneling time in (\ref{uncertain}), after
replacing $k_B T$ values from Eq.(\ref{temperature}), takes the
general form
\begin{equation}
\label{min-time}
\begin{split}
\Delta t = \frac{\hbar}{2 p_{t} \left( 2 \pi k_B T \right) } =  - \frac{\tau_c}{2\pi p_t}  e^{-2 \Phi} \left(\frac{1}{1+2\Phi} + \log \frac{1}{1+2 \Phi}\right) \\
\end{split}
\end{equation}
valid for any particle and any smooth potential. The tunneling time
formula (\ref{min-time}), will be hereon called {\it entropic
tunneling time} (ETT) to distinguish it from other tunneling time
definitions like phase time Eq. (\ref{phase-rec-time}) and dwell
time Eq. (\ref{dwell-rec-time}), which exist in the literature.

Before closing, it proves complementary to discuss the nature of the
ETT. In the classification of Busch \cite{intrinsic1,intrinsic2}, it
is an intrinsic dynamical time. The reason is that it is controlled
by particle's under-barrier momentum $\wp(x)$, which is integrated
over the barrier region to form $\Phi$. However, one must pay
attention that the energy-time uncertainty relation in
(\ref{uncertain}) is defined between $\Delta t$ and the thermal
energy $\Delta E_{ther}$ in (\ref{deltaE}) (not the fluctuations in
the true energy $E$ of the particle, which is held constant during
the entire tunneling dynamics). In this sense, though it is an
intrinsic dynamical time, the ETT can be contrasted with external
time measurements (as in Sec. 4 below) as it does not necessitate
any fluctuations in $E$ under the barrier
\cite{intrinsic1,intrinsic2}.

\section{Comparing ETT with Other Times}
\label{compare}}
{There are already manifold time definitions in the literature
\cite{time-review,time-review2,time-review3,time-review4}. They vary
in their origins, formulations and predictions. Two of them, the
Larmor and Buttiker-Landauer times, are special in that they are
defined via not the potential $V(x)$ alone but with its purposive
modifications. The Larmor time is based on the Larmor precession of
spin when the classically forbidden region is covered by an external
magnetic field \cite{larmour, larmour4, larmour5}. The
Buttiker-Landauer time is defined via an oscillating barrier
\cite{buttiker-landauer, buttiker-landauer2, buttiker-landauer3}.
These two times have been argued \cite{japan-rec, olkhovsky} to be
not the means but deviations of the tunneling time distributions.
They have obviously nothing in common with the ETT, which is based
on the potential barrier $V(x)$ alone. On the other hand, two
well-known time definitions, the phase time
\cite{time-review,time-review2,time-review3,time-review4,wigner,wigner2,wigner3}
and the dwell time
\cite{time-review,time-review2,time-review3,time-review4,larmour},
which, just like the ETT, are based on the potential energy $V(x)$.
They do not involve any external agents like the magnetic field used
in Larmor time. These three time definitions: ETT, phase, and dwell,
having a common setup, can thus be directly contrasted to determine
their physical relevance. To this end, as a simple setup, one can consider a rectangular potential barrier. (One must, however, keep in mind that a rectangular barrier does not quite fit to the WKB criteria. Nevertheless, it provides a viable framework to compare the time definitions. In case of worry, it can be approximated through a steep $\tanh$ potential.) Then, for a barrier of height $V_0$ and width
$L$, one derives the tunneling transmission probability
\begin{eqnarray}
\label{simrecpot}
p_{t}^{\Box}=\frac{1}{1+\frac{V_0^2 \sinh^2\Phi}{4E(V_0 - E)}}
\end{eqnarray}
where $\Box$ symbolizes the rectangular potential. With this transmission probability, the ETT takes the form
\begin{eqnarray}
\label{min-time-srp}
\begin{aligned}
\left(\Delta t\right)_{ETT}^{\Box} =  - \frac{\tau_{c}^{\Box}}{2\pi (V_0 - E)}\left((V_0 - E)+\frac{(V_0\sinh\Phi^{\Box})^2}{4E}\right) e^{-2 \Phi^{\Box}} \\ \times \left(\frac{1}{1+2\Phi^{\Box}} + \log \frac{1}{1+2 \Phi^{\Box}}\right)
\end{aligned}
\end{eqnarray}
in which
\begin{eqnarray}
\label{classictimesrp}
\Phi^{\Box}=\frac{\sqrt{2m(V_0-E)}}{\hbar}L
\end{eqnarray}
is the abbreviated action in the classically forbidden region as follows from (\ref{phase}), and
\begin{eqnarray}
\label{classtime-box}
 \tau_{c}^{\Box}=\frac{mL^2}{\hbar \Phi^{\Box}}
\end{eqnarray}
is the traversing time $-i\tau^{\Box}$ of the particle obeying
classical motion laws as follows from its definition in
(\ref{dc-time}), and $L=x_{R}-x_{L}$ is the barrier width.}

{The phase time (designated by $\varphi$), deriving from the
stationarity of the phase of the transmission amplitude
\cite{time-review,time-review2,time-review3,time-review4,wigner,wigner2,wigner3},
takes the form
\begin{eqnarray}
\label{phase-rec-time}
\left(\Delta t\right)^{\Box}_{\varphi} &=&\frac{(\tau_{c}^{\Box}) (p_{t}^{\Box})}{2 (\Phi^{\Box})^2 \Phi_E^3} \Big[\Phi^{\Box}\Phi_E^2 \left( (\Phi^{\Box})^2-\Phi_E^2\right)\nonumber\\
&+&\left( (\Phi^{\Box})^2+\Phi_E^2\right)^2\sinh\Phi^{\Box} \cosh\Phi^{\Box} \Big]
\end{eqnarray}
where $\Phi_E=\sqrt{2mE L^2}/\hbar$. In this phase time, the
hyperbolic functions arise from the phase of the transmission
amplitude.}

{The dwell time, expressing how long the particle stays in the
barrier region
\cite{time-review,time-review2,time-review3,time-review4,larmour},
is given by
\begin{eqnarray}
\label{dwell-rec-time}
\left(\Delta t\right)^{\Box}_{D} &=&\frac{(\tau_{c}^{\Box})  (p_{t}^{\Box})}{2 (\Phi^{\Box})^2 \Phi_E} \Big[\Phi^{\Box}\left((\Phi^{\Box})^2-\Phi_E^2\right)\nonumber\\
&+&\left((\Phi^{\Box})^2+\Phi_E^2\right)\sinh\Phi^{\Box} \cosh\Phi^{\Box} \Big]
\end{eqnarray}
where, compared to the phase time, lower powers of $\Phi^{\Box}$ and $\Phi_E$ are involved.}

{The three tunneling times the ETT (denoted by $\left(\Delta
t\right)_{ETT}^{\Box}$), the phase time (denoted by $\left(\Delta
t\right)_{\varphi}^{\Box} $) and the dwell time (denoted by
$\left(\Delta t\right)_{D}^{\Box} $) are seen to be distinct
functions. They lead thus to different predictions for time spent
during tunneling.  Nevertheless, to reveal their physical relevance
it is convenient to compare them for wide potentials ($L \rightarrow
\infty$):
\begin{eqnarray}
\label{unphys}
\left(\Delta t\right)_{ETT}^{\Box}&\rightarrow& \infty\\ \left(\Delta t\right)^{\Box}_{\varphi} &\rightarrow& \frac{\hbar}{E} \sqrt{\frac{E}{V_0-E}}\\
\left(\Delta t\right)^{\Box}_{D}&\rightarrow& \frac{\hbar }{V_0} \sqrt{\frac{E}{V_0-E}}
\end{eqnarray}
where the ETT is seen to diverge as expected of a potential barrier
of infinite width. The phase and dwell times, however, give the
unphysical result that it takes a finite time to traverse an
infinitely wide potential barrier. These two times suffer from
superluminality. More strikingly, those finite times vanish as
$V_{0} \rightarrow \infty$, meaning that the particle tunnels
through an infinitely wide and infinitely high potential barrier
instantaneously. This effect, the Hartman effect \cite{hartman},
renders the phase and dwell times unphysical. Needless to say, the
ETT is subluminal and suffers from no unphysical aspects like the
Hartman effect. {Moreover, relationship of tunneling time to
particle's dynamical transport has been verified for electrons in
\cite{verified} and discussed by Kullie in \cite{discussed}.}}

{Furthermore, apart from tunneling time definitions above, the
complex tunneling times are hard to make sense \cite{complex}. The
path integral averages of the classical time
\cite{times-are-averages,times-are-averages2,times-are-averages3}
and of the Larmor time \cite{leavens-and-aers,leavens-and-aers2}
give rise to complex times. They also arise via scattering-theoretic
formulation \cite{pollak-miller}. Their real and imaginary parts are
related to other tunneling times in specific ways
\cite{time-review,time-review2,time-review3,time-review4,japan-rec}.
Unlike them, the ETT is purely real and bears no relation to complex
times.}

\section{Confronting ETT with Experiment}
In this section, we shall perform numerical analysis to test the ETT against certain experimental results.
\subsection{Laser-Driven He Ionization}
\begin{figure}
\begin{center}
\epsfxsize=4in
\epsfbox{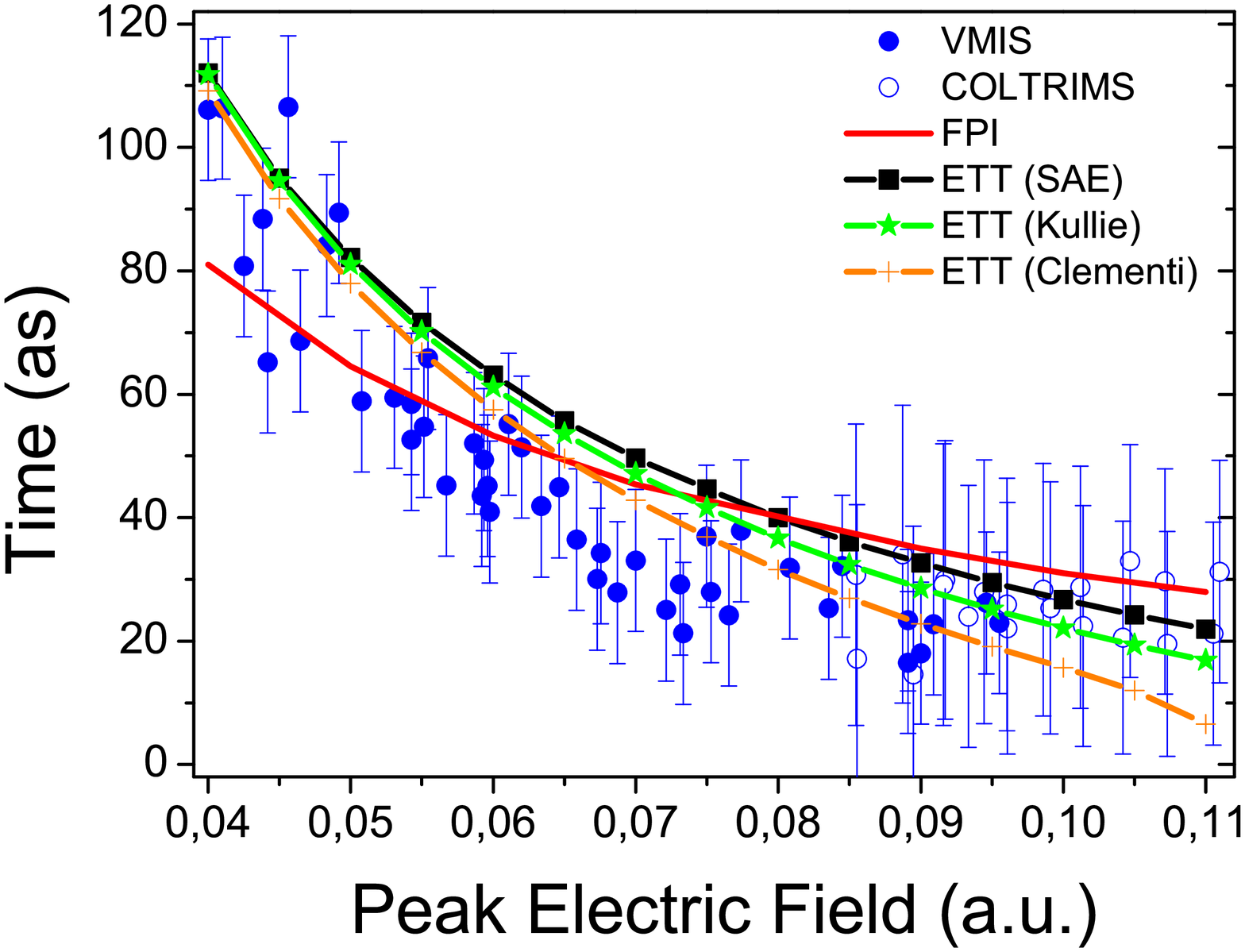}
\epsfxsize=4in
\epsfbox{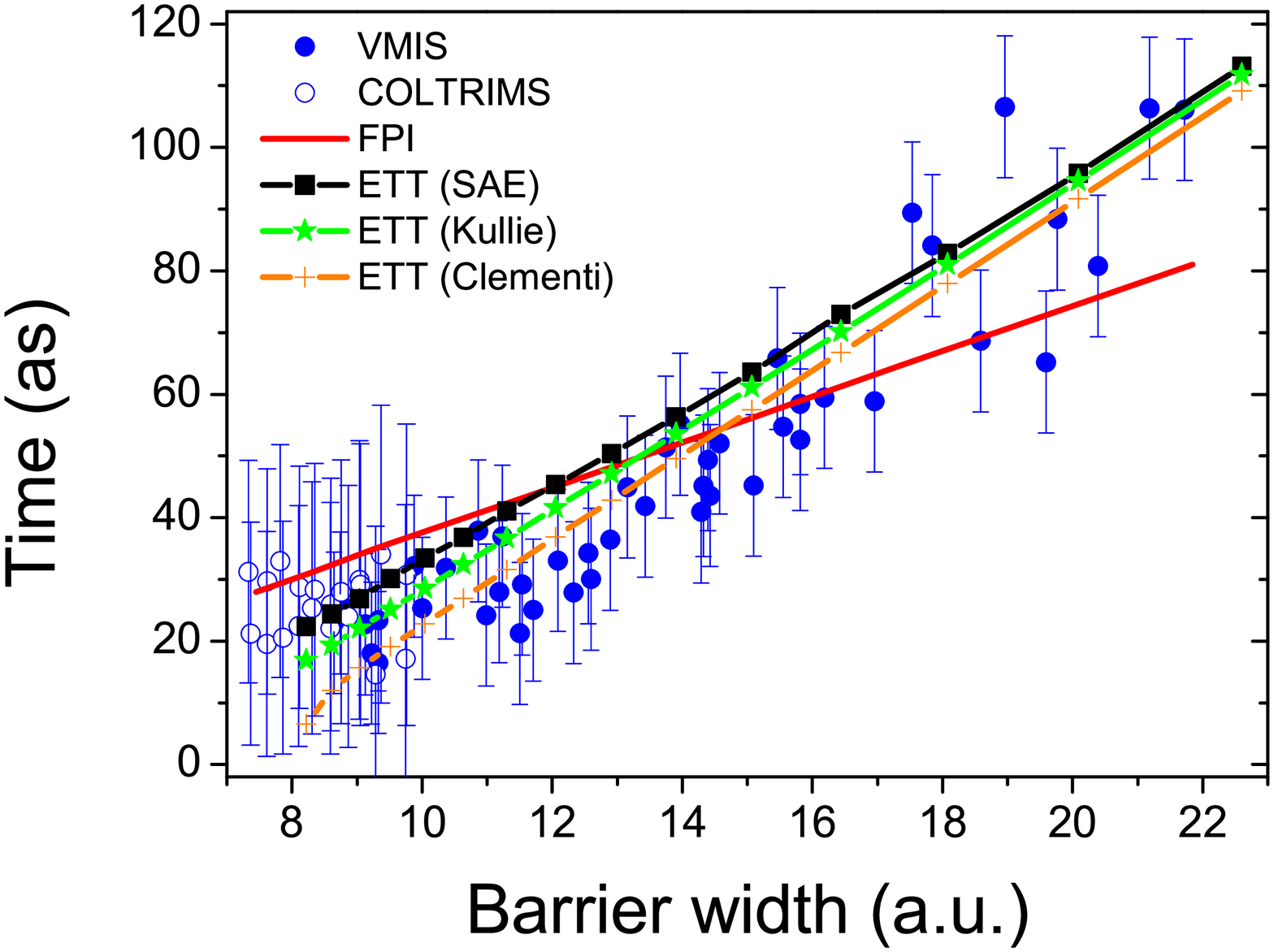}
\caption{Tunneling times as functions of peak electric field ${\mathcal{E}}$ (top panel) and
experimental barrier width ($E/{\mathcal{E}}$) (bottom panel). {The entropic
tunneling time (ETT) is depicted by filled square, star and plus representing SAE, Kullie and Clementi \textit{et al} models, respectively.} It is seen to agree with the experimental
data throughout. Both panels explicitly show how the entropic time adheres to the experimental
data \cite{exp-incele}) while the FPI time diverges away from data at the asymptotics.}
\label{figure}
\end{center}
\end{figure}

Electric fields of high-intensity lasers reshape Coulomb potential
in atoms to form a potential barrier through which electrons can
tunnel to continuum \cite{keldysh,keldysh2}. At the peak value
${\mathcal{E}}$ of the electric field, one of the electrons in He
possesses {the effective potential energy
\begin{eqnarray}
\label{pot}
V(x) = - \frac{Z_{eff}}{x} - {\mathcal{E}} x\;\;\;\;\;\;
\end{eqnarray}
which is parametrized in terms of  an effective nuclear charge
$Z_{eff}$.} The transmission probability takes the form
\cite{trans,ttez}
\begin{eqnarray}
\label{transcoef}
p_{t}^{He} = \frac{1}{\cosh^2\Phi}
\end{eqnarray}
{after solving (\ref{en-eq}) and matching the solutions at each of
the turning points $x_{L,R}$. With this transmission probability,
ETT (\ref{min-time}) becomes
\begin{equation}
\label{min-time-wkb}
\begin{split}
(\Delta t)^{He} =  - \frac{\tau_c}{2\pi} \cosh^2\Phi\, e^{-2 \Phi} \left(\frac{1}{1+2\Phi} + \log \frac{1}{1+2 \Phi}\right) \\
\end{split}
\end{equation}
On the other hand, advancements in ultrafast science, where strong
laser fields are used to ionize atoms by quantum tunneling, are
capable of observing tunneling transition and measuring the
tunneling time \cite{exp0,exp1,exp11,exp12,exp13,exp2,exp2-new}. In
spite of various factors affecting the experiments
\cite{exp-problem1,exp-problem2, exp-resolve1, exp-resolve2},
improving on previous single-particle tunneling time measurements
\cite{exp2} by using attoclock in strong laser fields
\cite{exp2-new}, in 2013 {the research team of Ursula Keller at ETH
Zurich} have performed a refined measurement of the tunneling time
of electrons in He atom \cite{exp-incele}. Moreover, their
measurements have been shown to be stable \cite{yeni-deney} (see
also the simulation study \cite{yeni-deney2}) against non-adiabatic
effects \cite{exp-problem1, exp-problem2}. As a result, using the effective potential (\ref{pot}) enables us to use this
experiment as a testbed for the ETT. The limiting factor here is validity of the WKB solution. The WKB approximation holds good for smooth potentials. The same is not true for steep potentials with sharp edges. For such potentials, beyond-the-WKB effects
can be significant. For the He ionization problem at hand, the WKB approximation hardly works at large laser field strength ${\mathcal{E}}$ for which potential is steep (see the potentials in \cite{triangular}). In such regions, agreement with experiment can require beyond-the-WKB effects to be incorporated. Besides, one keeps in mind that the laser field, depending on its strength, can cause either multiphoton or tunneling ionization. And a sensible comparison of the ETT with experiments is possible only in tunneling regime
corresponding to the Keldysh parameter \cite{keldysh,keldysh2} range
$\gamma \lesssim 1$.

In the experimental setup, laser intensity $3.478\times 10^{16}\
{\rm W}/{\rm cm^2} {\mathcal{E}}^2$ is varied from $0.730 \times
10^{14}\ {\rm W}/{\rm cm^2}$ to $7.50 \times 10^{14}\ {\rm W}/{\rm
cm^2}$ by varying the peak electric field ${\mathcal{E}}$ from
$0.04$ to $0.11$ in atomic units. The electron energy $E = - 0.904\
{\rm a. u.}$ is the first ionization {potential} of the He atom.
Momentum distribution of the liberated electrons are obtained by
cold-target recoil-ion momentum spectrometer (COLTRIMS) and by
velocity map imaging spectrometer (VMIS) (see the experiment section
of \cite{exp-incele} for details). The VMIS is used particularly at
low laser intensities. Quantum tunneling is ensured to be the
dominant ionization mechanism by keeping the Keldysh parameter small
($\gamma \lesssim 1$). The experimental results are given in Fig. 3
of \cite{exp-incele}. Depicted in Fig. 3 (a) {of \cite{exp-incele}}
are different tunneling times \cite{buttiker-landauer, larmour,
pollak-miller,wigner,times-are-averages,times-are-averages2,times-are-averages3}
contrasted with experiment's own results. {Similarly, given in Fig.
3 (b) and (d) of \cite{exp-incele}}, are tunneling times as
functions of the peak electric field ${\mathcal{E}}$ and barrier
width (approximated as $E/{\mathcal{E}}\ {\rm a. u.} $ in the
experiment). The experiment (as well as \cite{yeni-deney}) also
indicates that among all widely-used tunneling time definitions only
the FPI time comes closest to its measurements (see Fig. 3 (b) and
(d) {of \cite{exp-incele}}). {Additionally, \cite{overestimate} also
indicates that the phase and dwell times both overestimate the
experimental result \cite{exp-incele}.}

For confronting the ETT (\ref{min-time-wkb}) with experiment, it
suffices to replace the potential energy (\ref{pot}) in $\Phi$, and
$\tau_c$, and evaluate them with the turning points $x_{L}$ and
$x_{R}>x_L$ satisfying
\begin{eqnarray}
\label{roots}
x_{L(R)}=\frac{E - (+) \sqrt{E^2 - 4Z_{eff}(x_{L(R)}){\mathcal{E}}}} {2{\mathcal{E}}}
\end{eqnarray}
which are the roots of the vanishing kinetic energy condition
$V(x)-E=0$. It is clear that better the knowledge of $Z_{eff}$
better the prediction of tunneling time (through the turning points
and hence $\Phi$). Following the literature, three different
$Z_{eff}$:
\begin{enumerate}
\item SAE potential \cite{SAE}
\begin{eqnarray}
Z_{eff}=Z+a_1 e^{-a_2 x}+a_3 x e^{-a_4 x}+a_5 e^{-a_6 x}
\end{eqnarray}
in which $Z=1$, $a_1 = 1.231$, $a_2 = 0.662$, $a_3 = -1.325$,
$a_4 = 1.236$, $a_5 = -0.231$, and $a_6 = 0.480$ in atomic
units.
\item Kullie constant \cite{uncertain,kullie} $Z_{eff} = 1.375$, and
\item Clementi \textit{et al} constant \cite{clementi} $Z_{eff} = 1.6875$.
\end{enumerate}
were used here. The turning points in (\ref{roots}) are directly evaluated for
constant $Z_{eff}$. For the SAE potential, however, it is necessary
to find a self-consistent solution for $x_{L,R}$. This is done by
starting with a random $x$ value and iterating it $N$ times until
$V(x_N)-E< 10^{-4}$. We list down $x_L$ and $x_R$ values and the
corresponding classical time $\tau_c$ and the ETT in the Table
\ref{tab} by considering two peak values for the laser field.
The ETT for different $Z_{eff}$ are plotted in
Fig. \ref{figure} as functions of the peak electric field
${\mathcal{E}}$ and experiment's barrier width $E/{\mathcal{E}}$
(which is not the true barrier width $x_R - x_L$). It is
superimposed on Fig. 3 (b) and (d) of \cite{exp-incele}.} The figure
manifestly shows that ETT exhibits good agreement with the
experimental data for all three potential models. They, nevertheless, start
differing at large laser field ${\mathcal{E}}$ values. This can be understood as follows:
\begin{enumerate}
\item At small $\mathcal{E}$, the potential $V(x)$, away from the nucleus, is smooth in all three cases (SAE, Kullie and Clementi). It supports the WKB solution. This is confirmed by 
the fact that ETT(SAE), ETT(Kullie) and ETT(Clementi) fairly agree at low laser field values $\mathcal{E}$ (see Fig. \ref{figure} below  $\mathcal{E}\simeq 0.08$). 

\item At large $\mathcal{E}$, the three times start differing significantly. The reason for this
is that potential $V(x)$ is no longer smooth, and depending on the potential parameters the ETT values diverge. The potential takes a right-triangular shape with smaller and smaller depression angle for larger and larger $\mathcal{E}$. The WKB solution hardly works for this potential (see the detailed analysis in \cite{triangular}). To have a quantitative understanding of the ETT curves at large $\mathcal{E}$, it suffices to consider a right-triangle potential $V(x) = V_0 - \mathcal{E} x$ extending from $x=0$ to $L$ (having the same width as the rectangular potential in Sec. 3) for which one gets
\begin{eqnarray}
\Phi^{\triangle} = \frac{2}{3} \frac{(V_0-E)}{{\mathcal{E}} L} \Phi^{\Box}
\end{eqnarray}
and
\begin{eqnarray}
\label{class-time}
\tau_c^{\triangle} = \frac{2(V_0-E)}{{\mathcal{E}} L} \tau_c^{\Box}
\end{eqnarray}
so that  $p_m^{\triangle}>p_m^{\Box}$ and $p_t^{\triangle}>p_t^{\Box}$. Using these in the 
general ETT formula in (\ref{min-time}) one can compute tunneling time for the triangular potential. It is clear that $\Phi^{\triangle}< \Phi^{\Box}$ and $\tau_c^{\triangle}< \tau_c^{\Box}$ under sufficiently strong laser fields {\it i.e.} large $\mathcal{E}$. This means that the ETT curves in Fig. \ref{figure} will be pushed down depending on the laser field strength. In fact, the classical time (\ref{class-time}) is already sufficient to understand this. The simple triangular potential, though different than the actual potential in (\ref{pot}), is powerful enough to reveal the essential features of the ETT curves in Fig. \ref{figure}. 

All might seem fine, but one must still keep in mind that the triangular potential (like the He ionization at large $\mathcal{E}$) is sensitive to beyond-the-WKB effects. The WKB-based ETT predictions may not therefore be accurate. Indeed, the ETT(Clementi), for instance, takes smaller values than expected (though $Z_{eff}=1.6875$ is reasonable in the large $\mathcal{E}$ domain). This can be understood as an artifact of the beyond-the-WKB effects. This problem, as discussed also in detail in \cite{triangular}, is a characteristic feature of  large $\mathcal{E}$ region.
\end{enumerate}
These properties help interpreting the ETT curves in Fig. \ref{figure} as a transition from smooth to steep potential regions. 

Out of the known time definitions, as seen from Fig.
\ref{figure}, only the FPI time \cite{times-are-averages,
times-are-averages2, times-are-averages3, FPI-method} seems to 
come closest to the experimental data. This congruence 
is, actually, somewhat biased by the way the probability 
distribution is coarse-grained in \cite{exp-incele}. 
(This point is discussed in detail in \cite{Kullie-new}.)
Besides this,  in the top panel, its predictions diverge from the data as the peak electric field
increases. In the bottom panel, it matches with the COLTRIMS data at
low barrier widths while it diverges at larger barrier widths. In
contrast to these divergent behaviors in the FPI time, ETT stays
congruent to experimental data for a fairly wide range of potential
parameters.

As a result, it would not be unrealistic to conclude that the ETT
shows good agreement with experiment, and furthermore, {outperforms the widely-used tunneling time models among those shown in Fig. 3 (a)
of \cite{exp-incele} and the FPI time.}

\begin{table}
\caption{\label{tab}Contrasting the three $Z_{eff}$ models, after setting ${\mathcal{E}}=0.04 a.u.$ $({\mathcal{E}}=0.11 a.u.$), in terms of their predictions for the turning points $x_{L,R}$, classical time $\tau_c$ and the ETT.}	
\begin{tabular}{lllll}
  &$x_L$ ($a.u.$) &$x_R$ ($a.u.$) &$\tau_c$ ($as$)&ETT ($as$)\\
   \hline
   SAE&1.24(1.39)&21.43(6.90)&833.82(312.24)&113.08(22.20)\\
   Kullie&1.64(2.02)&20.96(6.20)&850.73(322.72)&111.75(16.85)\\
   Clementi&2.05(2.87)&20.55(5.35)&856.49(326.50)&109.14(6.54)
\end{tabular}
\end{table}

\subsection{Electron Transfer Reactions}
Rectangular potential barriers, apart from their direct solubility,
prove useful in modeling tunneling systems whose potential barriers
are nearly constant. {Even though WKB approach does not work properly for this potential barrier shape, as already mentioned, one can still expected to obtain corroborative results.} As an example, efficient tunneling barrier
($\Delta E_{Eff}=V_0 -E$) proves to be a useful approach in the case
of long-range electron transfer reactions\cite{lrelect}. The
transmission probability and ETT of this simple case is given in
Eq.(\ref{simrecpot}) and Eq.(\ref{min-time-srp}) respectively.
Notice that, the exact transmission probability (\ref{simrecpot})
reduces to the WKB result in (\ref{transcoef}) only when $E =
V_0/2$.

In the case of electron transfer reactions, typical range for an
electron to transit from donor to acceptor is in between $5$(\AA)
and $30$(\AA) ( the range of electron transfer in proteins from
$15$(\AA) to $30$(\AA) \cite{etprotein}).
\begin{figure}
\begin{center}
\includegraphics[scale=0.35]{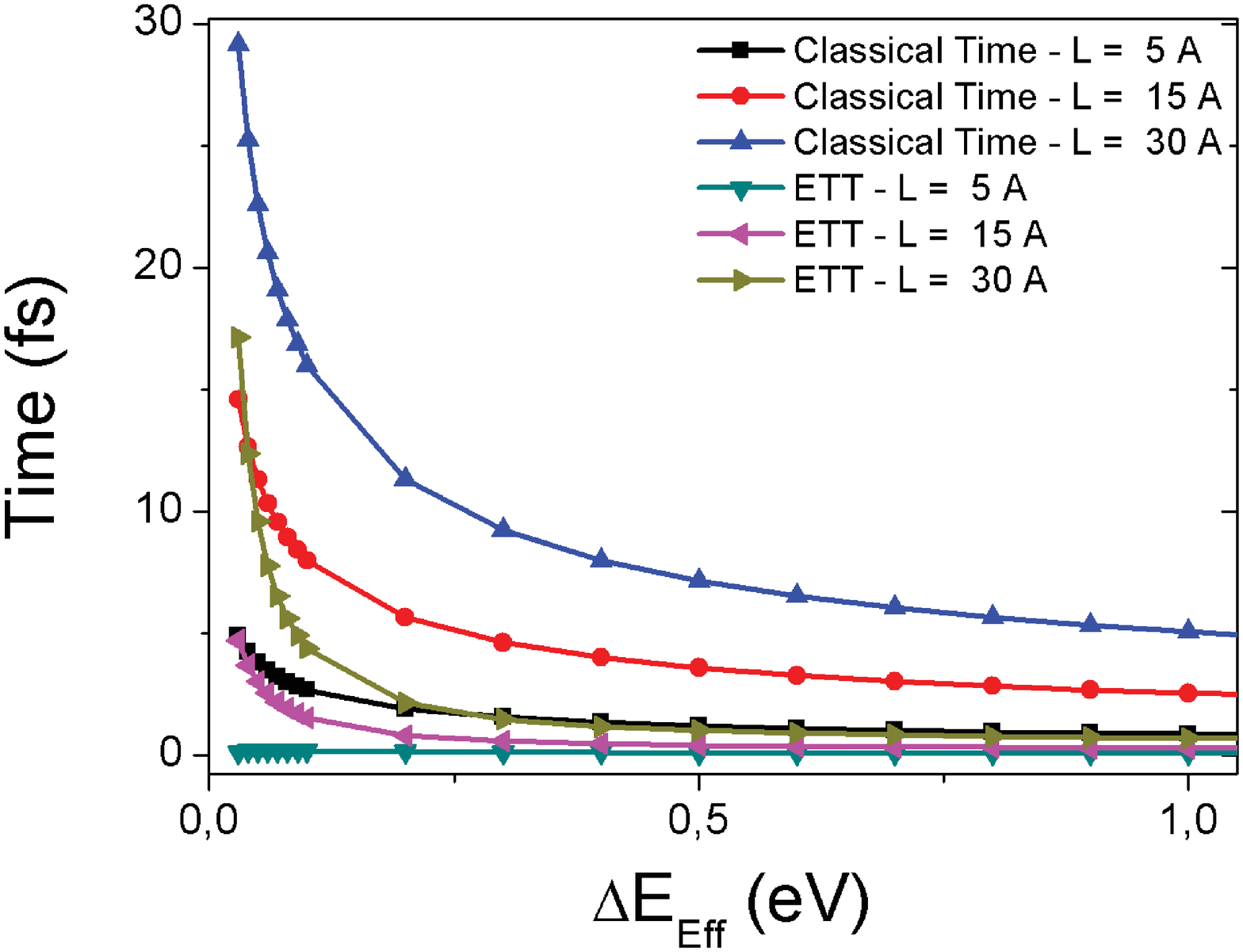}
\includegraphics[scale=0.35]{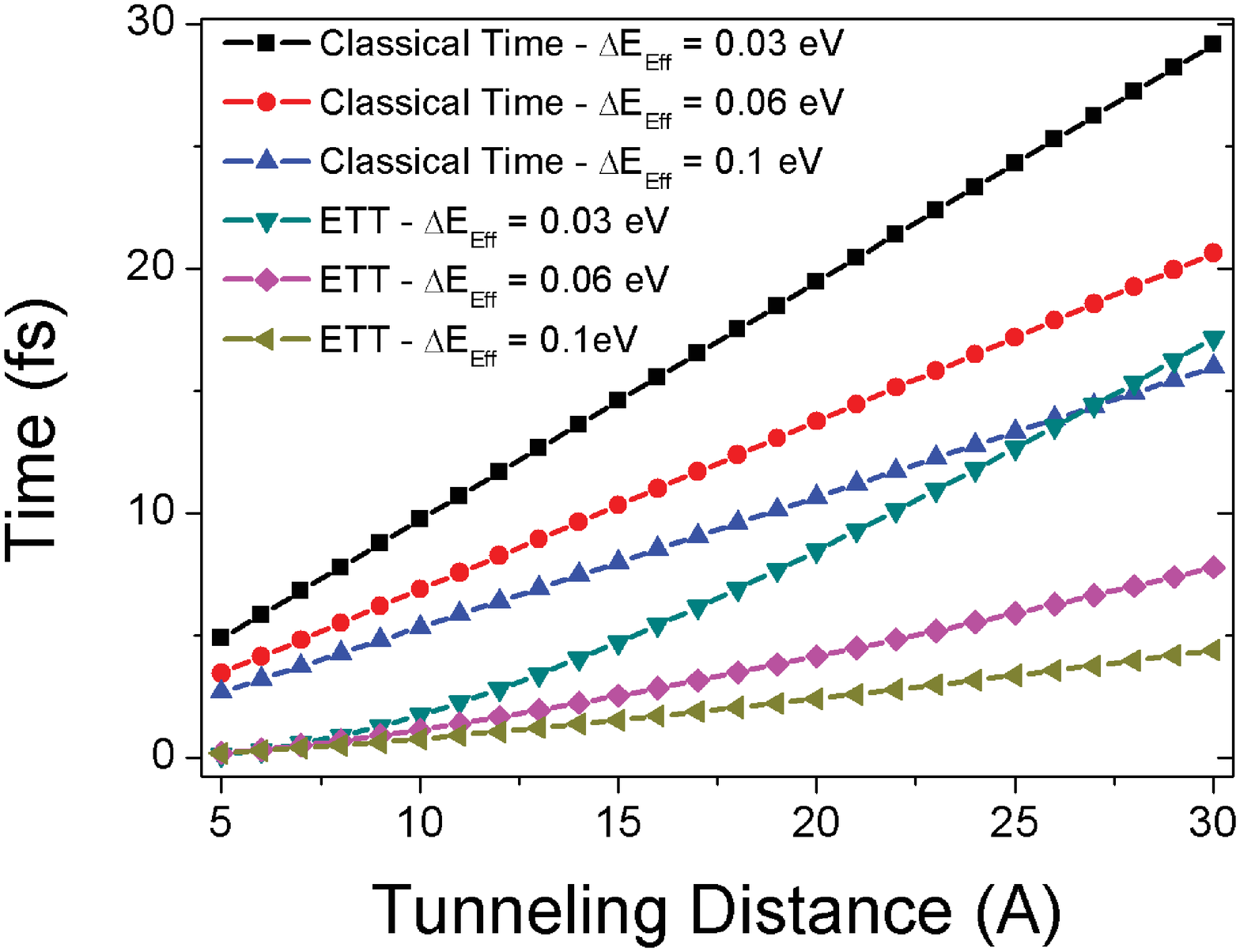}
\caption{{Electron transition time as functions of tunneling distance given for different $\Delta E_{Eff}$ values(top panel) and $\Delta E_{Eff}$ given for different tunneling distance values(bottom panel) calculated in terms of both classical time and ETT. Energy of an electron $E$ is taken as $1 eV$.}}
\label{figure2}
\end{center}
\end{figure}
Nuclear vibrations have typically $<3000$ $\rm{cm^{-1}}$ frequency
range. Corresponding half-period of vibrations are thus above the
$5$ femtosecond ($fs$) scale. Using ETT and classical time in
(\ref{classictimesrp}), electron transition times can be calculated.
It is found that, classical time is comparable to the nuclear
vibrations in Fig.(\ref{figure2}). On the other hand, ETT gives
lower values compared to classical time, in both cases. From this,
one recalls that $\tau_c$ is the absolute value of the imaginary
time spent during the classical motion. The difference between the
two tunneling times indicates how strong the quantum effects are.
Furthermore, it is seen that these values of tunneling time are even
lower than half period of the corresponding nuclear vibrations,
whereby proving the predomination of the adiabatic
region\cite{electrontransfer}. However, it is for the $\Delta
E_{Eff}<0.1$ $eV$ and the tunneling distance $>15$ (\AA) that the
tunneling time becomes comparable to the time scale of nuclear
vibrations. Time scale compatibility between nuclear vibrations and
tunneling time allows overlap of molecular configuration change with
the electron transition process while electron is tunneling. Due to
energy conservation, configuration change during the electron
tunneling leads to an energy exchange between electron and nuclei,
which entangles them. Therefore, in this restricted region with time
scale comparability, it becomes possible for an electron to make
transition from BO to non-BO regime where acceptor-donor
wavefunctions collapse. Similar analysis was performed already in
\cite{BOapprox}, where their tunneling time definition for simple
rectangular potential barrier equals to the classical time
definition in (\ref{classictimesrp}). It follows the same trend with
classical time given in Fig.\ref{figure2}. It obviously lacks
quantum contributions and proves thus insufficient for having a
clear picture of BO to non-BO transition regime. As a result, the
conditions determined by ETT set the range of energy difference and
tunneling distance more accurately for long range electron transfer
reactions to take place.
\section{Conclusion and Future Prospects}
ETT, with its statistical conception, subluminal nature and
experimental confirmation, works through as a realistic model of the
tunneling time. It provides a quantum theoretic framework by which
one can analyse all kinds of tunneling-enabled phenomena ranging
from STM to DNA mutation and flash memory to interstellar chemistry.
This rather widespread role facilitates phenomenological tests and
possible improvements of the entropic formalism through variety of
sources.

The tunnel effect, a manifestation of the evanescent wave behaviour, can occur in all wave phenomena.
A generic wave equation
\begin{eqnarray}
\label{wave-eq}
\frac{d^2}{dx^2} W(x) = -k^2(x) W(x)
\end{eqnarray}
portrays a propagating wave for $k(x)\in \Re$ and evanescing wave for $k(x)=i \kappa(x) \in \Im$.
Pragmatically, ETT formalism can be extended to this wave behaviour
with the identification
\begin{eqnarray}
\frac{\wp(x)}{\hbar} \rightarrow \kappa(x)
\end{eqnarray}
as revealed by contrasting (\ref{wave-eq}) with (\ref{en-eq}). To
make sense of this formal equivalence, it is necessary to determine
first the origin of the imaginary, inhomogeneous wavenumber
$\kappa(x)$ in view of (\ref{c-mom}). Indeed, monochromatic wave
must have a frequency below the natural cut-off frequency of the
medium for evanescent behaviour to occur. Next, it is necessary to
construct the quanta corresponding to the wave so that evanescing
characterizes the tunneling phenomenon. Finally, it is necessary to
establish an analogy with the Schroedinger equation by taking into
account the symmetries of the wave equation.

There are numerous wave phenomena. The probability waves of quantum
theory, $W(x) \equiv \psi_e(x)$, govern electron tunneling in
semiconductors, Hydrogen tunneling in biochemical systems and Helium
tunneling in nuclear systems. The electric waves, $W(x) \equiv
{\mathcal{E}}(x)$, describe photon tunneling in materials with
imaginary refractive index (band gaps, dielectric gaps, air gaps)
\cite{ott,ott2,ott3}. The photonic STM \cite{pstm,pstm2,pstm3},
scanning of surfaces with a fiber optic tip, is a direct application
of photon tunneling. The sound waves, on the other hand, encode
phonon tunneling through acoustic band gaps
\cite{sound,sound2,sound3,sound4}. Tunneling of the thermal
vibrations of an STM tip to the sample is a direct realization of
the phonon tunneling \cite{yeni-prl}. The optical and acoustic
tunneling studies have been thoroughly reviewed in \cite{emt}
experiment by experiment. The photon and phonon tunneling processes,
interpreted so far only with phase and dwell times \cite{emt}, need
be analysed and reinterpreted within ETT formalism, as is being
planned to be done in upcoming work.

The ETT is new. It is theoretically consistent and experimentally
pertinent for smooth potentials. In fact, one direction to improve on
the present work would be to include of subleading WKB corrections 
(involving derivatives of the potential). Besides, the ETT can be 
tested against experimental data on appropriate physical, chemical and
biological processes. 

\section{Acknowledgments}
The authors thank to Onur Rauf Y{\i}lmaz for reading the manuscript.
They are grateful to Dr. Ossama Kullie for useful e-mail exchange
and enlightening comments. {They thank to A. Landsman (through C.
Hofmann) for the experimental data used in Fig. 3.} The authors
thank to contentious referee for constructive suggestions.


\begin{thebibliography}{99}
\bibitem{neden}
It is possible to consider alternative forms for entropy such as $k_B p_m \log
  (p_m - \log p_m)$. This form, however, though vanishes at $p_m=0$ and $p_m=1$
  just as the entropy (\ref{entropy}), necessitates $e^{- 2 \Phi} + 2\Phi$
  number of microstates, not the right amount of $1 + 2\Phi$.
\end{thebibliography}
\end{document}